\documentclass{PoS}

\usepackage{epsfig}
\usepackage{psfrag}
\usepackage{amsmath}

\newcommand{\beqa}{\begin{eqnarray}}
\newcommand{\eeqa}{\end{eqnarray}}
\newcommand{\beq}{\begin{equation}}
\newcommand{\eeq}{\end{equation}}
\newcommand{\p}{\partial}
\newcommand{\pslash}{\p \hspace{-6pt}\slash\hspace{+1pt}}
\newcommand{\mvcapt}{\vspace{-0.5cm}}

\rightline{\sffamily \small CERN-PH-TH/2006-214}

\title{New baryon matter in the lattice Gross-Neveu model}

\ShortTitle{New baryon matter in the GN model}
\author{Philippe de Forcrand$^{a,b}$ and \speaker{Urs Wenger}$^{ b}$\\
        \llap{$^a$}CERN, Physics Department, TH Unit,
        CH--1211 Geneva 23, Switzerland\\
        \llap{$^b$}ETH Zurich, Institute for Theoretical Physics,
        CH--8093 Zurich, Switzerland\\
        E-mail: \email{forcrand@phys.ethz.ch}, \email{wenger@phys.ethz.ch}}
     
\abstract{ 
  We investigate the Gross-Neveu model on the lattice at finite temperature
  and chemical potential in the limit of an infinite number of fermion
  flavours.  We check the universality of the continuum limit of staggered and
  overlap fermions at finite temperature and chemical potential.  We show that
  at finite density a recently discovered phase of cold baryonic matter
  emerges as a baryon crystal from a spatially inhomogeneous fermion
  condensate. However, we also demonstrate that on the lattice, this new phase
  disappears at large coupling or in small volumes. Furthermore, we
  investigate unusual finite size effects that appear at finite chemical
  potential.  Finally, we speculate on the implications of our findings for
  QCD.
}

\FullConference{XXIV International Symposium on Lattice Field Theory\\
                 July 23-28 2006\\
                 Tucson Arizona, US}

\begin{document}

\section{Introduction}
In order to understand the phases of matter at finite temperature and density
it is necessary to understand the properties of the non-perturbative vacuum of
QCD or related models. The Gross-Neveu model \cite{Gross:1974jv} resembles QCD
in many respects and can be solved analytically in the limit of an infinite
number of fermion flavours $N$. Here we report on our investigation of this
model on a discrete space-time lattice at finite temperature and density in
the large-$N$ limit.

\section{The Gross-Neveu model}
Let us start with writing the Euclidean Lagrangian density of the Gross-Neveu
(GN) model \cite{Gross:1974jv} in $1+1$ dimensions,
\beq 
{\cal L} =
\sum_{\alpha=1}^{N} \bar \psi^\alpha(x) \pslash \psi^\alpha(x) - \frac{g^2}{2}
\left( \sum_{\alpha=1}^{N} \bar \psi^\alpha(x) \psi^\alpha(x) \right)^2, 
\eeq
where $\psi^\alpha(x)$ are 2-component Dirac spinors and $\alpha$ is a flavour
index. Usually, one introduces a scalar field $\sigma(x)$ conjugate to
$\sum_{\alpha=1}^{N} \bar \psi^\alpha(x) \psi^\alpha(x)$ in order to transform
away the 4-fermion coupling,
\beq
\label{eq:bilinear_GN-model}
{\cal L} = \sum_{\alpha=1}^{N} \bar \psi^\alpha(x) \pslash
  \psi^\alpha(x) + 
  \frac{1}{2 g^2} \sigma(x)^2 + \sigma(x) \sum_{\alpha=1}^{N}
  \bar \psi^\alpha(x) \psi^\alpha(x).
\eeq  

The GN model shares many interesting properties with QCD. In particular, it is
renormalisable and asymptotically free, with the $\beta$-function given to
lowest order by $\beta(g) = - \frac{N-1}{2 \pi} g^3 + O(g^5)$.  Moreover it
enjoys a $O(2N) \times \Gamma$ global symmetry, where $\Gamma$ is the discrete
chiral symmetry
\beq 
\Gamma: \qquad \psi \rightarrow \gamma_5 \psi,
\quad \bar \psi \rightarrow -\bar\psi \gamma_5, \quad \sigma \rightarrow -
\sigma, 
\eeq
and it exhibits spontaneous breaking of this discrete chiral symmetry which in
turn leads to the fermions acquiring a non-vanishing mass $\sigma_0 = \langle
\sigma \rangle$ (dimensional transmutation)\footnote
{Note that there is no Goldstone boson since $\Gamma$ is a discrete
  symmetry.}.

In the large-$N$ limit where the number of fermion flavours $N$ is taken to
infinity while keeping $\lambda = g^2 N$ fixed, the model can be solved
analytically. One can integrate out the fermions to obtain $Z = \int
{[d\sigma]} \exp\left\{ - S_{\textrm{\tiny eff}} \right\}$ with
\beq
 S_{\textrm{\tiny eff}} = N \left\{ \int {dx} \frac{\sigma(x)^2}{2 \lambda} - 
      \textrm{Tr}
  \log \left[ \pslash + \sigma \right] \right\}.
\eeq
The minimum of the effective potential is given by a set of equations,
\beq
\label{eq:self_consistency_equation}
\p_{\sigma(x)} S_{\textrm{\tiny eff}}/N = \frac{\sigma(x)}{\lambda} -
\p_{\sigma(x)}  \textrm{Tr}
  \log \left[ \pslash + \sigma \right] = 0, \quad \forall x,
\eeq
and for a homogeneous condensate $\sigma(x)=\sigma$ this set reduces to a
single equation
\beq
\frac{\sigma}{\lambda} = 
\p_{\sigma}  \textrm{Tr}
  \log \left[ \pslash + \sigma \right],
\eeq
or in momentum space
\beq
\sigma=0 \quad \text{or} \quad \frac{1}{\lambda} = \int \frac{d^2k}{(2 \pi)^2}  \frac{2}{k^2 + \sigma^2}
\eeq
which is a self consistency equation for the fermion condensate or simply the
{\it gap equation}. Equivalent equations can be derived via Hartree-Fock,
Schwinger-Dyson or Bethe-Salpeter approaches. With the help of these equations
one can derive the spectrum of the GN model \cite{Dashen:1975xh} -- it
contains fermions with mass $m=\sigma_0$, $n$-fermion bound states, and
baryons with mass $m_B = \sigma_0 \cdot \frac{2N}{\pi}$.

\begin{figure}
\begin{minipage}{0.5\textwidth}
\begin{psfrags}
\psfrag{MU}{\huge $\mu/\sigma_0$}
\psfrag{TEMPERATUR}{\huge $T/\sigma_0$}
\psfrag{PUNKTA}{}
\psfrag{TRANSINV}{\Huge $\sigma \neq 0$}
\psfrag{CHIRAL}{\Huge $\sigma = 0$}
\psfrag{PUNKTB}{\huge $P_L$}
\psfrag{TKRITISCH}{\Huge $\frac{e^{\mathrm{C}}}{\pi}$}
\psfrag{CRYSTAL}{\Huge crystal phase}
\psfrag{WURZEL}{\Huge $\frac{1}{\sqrt{2}}$}
\psfrag{ZWEIPI}{\Huge $\frac{2}{\pi}$}
\vskip-0.4cm
\epsfig{file=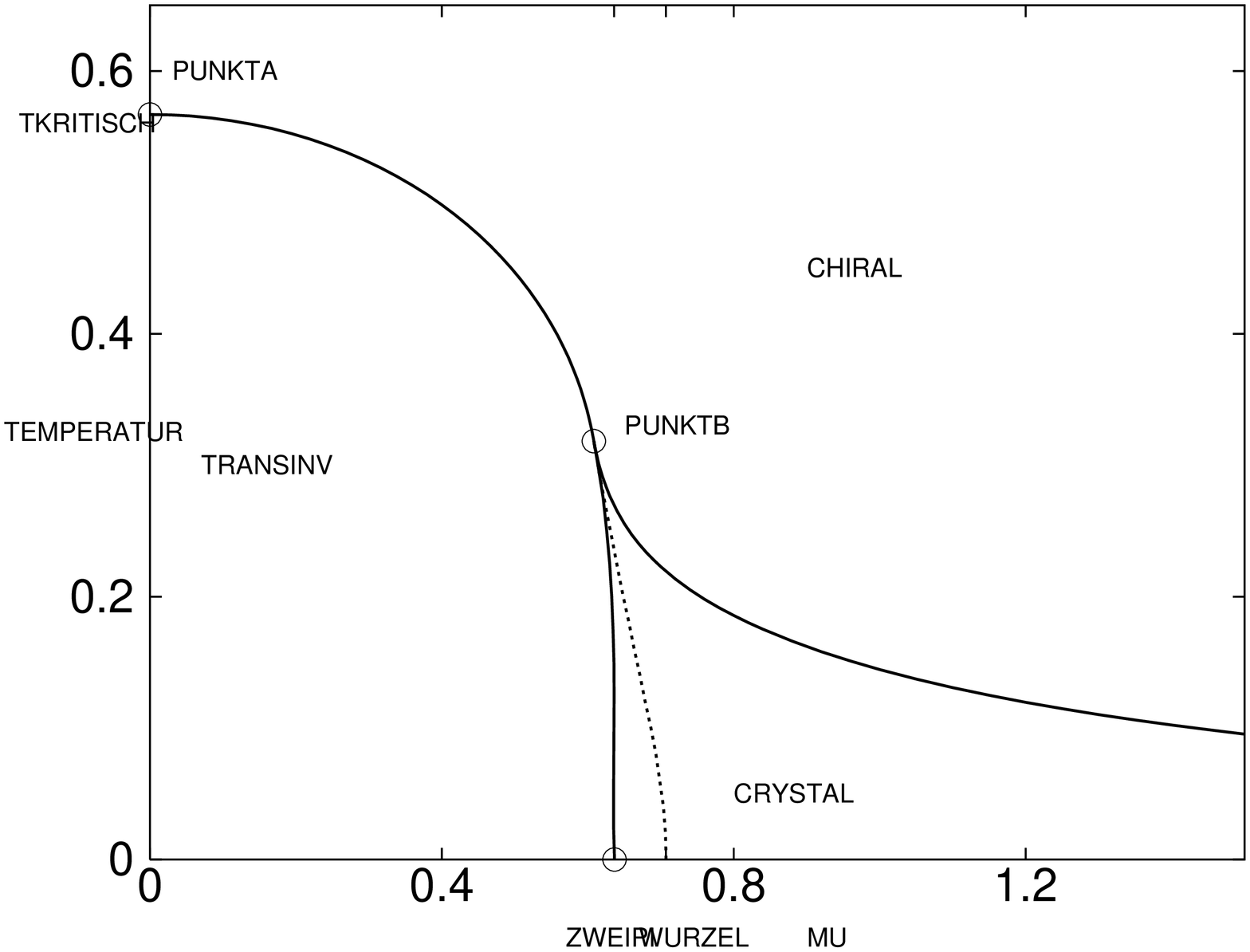, angle=-90, width=\textwidth}
\end{psfrags}
\end{minipage}
\begin{minipage}{0.5\textwidth}
\epsfig{file=Figures/crystal_phase_diagram_Lt80_Lx80_bw.eps,width=\textwidth}
\end{minipage}
\caption{
  Revised phase diagram of the GN model in the continuum from
  \cite{Thies:2003kk} (left) and on the lattice (right).  Full lines
  correspond to second order phase boundaries, dotted lines denote the
  (incorrect) first order phase boundary from the translationally invariant
  calculation.  $P_L$ is a multicritical Lifshitz point and $C$ is the Euler
  constant.  The dashed lines in the right figure illustrate the finite size
  artefacts, in particular the incommensurability effects at the right phase
  boundary of the crystal phase.
}
\vspace{-0.2cm}
\label{fig:revised_phase_diagram}
\end{figure}

What makes the GN model particularly interesting for our purpose is its rich
phase structure in the $(\mu,T)$--plane where $\mu$ represents the fermion
chemical potential and $T$ the temperature\footnote
{ In the large-$N$ limit, a discrete symmetry can break spontaneously in
  $(1+1)d$ even at finite temperature.  }.
In the homogeneous mean field approximation this was studied by Wolff
\cite{Wolff:1985av} who found two phases, a massive and a massless Fermi gas,
separated by a line of first and second order transitions meeting in a
tri-critical point, cf.~Figure \ref{fig:revised_phase_diagram}.  Recently the
phase structure has been further clarified by Thies and Urlichs
\cite{Thies:2003kk,Thies:2003br} who relaxed the tacit assumption of
translational invariance of the condensate. Motivated by the fact that matter
at low density forms isolated baryons they analytically found an inhomogeneous
solution $\sigma(x)$ to eq.(\ref{eq:self_consistency_equation}). Indeed, they
discovered that there exists a new baryonic matter phase at low temperature
where baryons form a crystal structure. The transitions from the massive to
the crystal phase and further to the massless phase are all second order.

\section{Lattice formulations of the Gross-Neveu model}
Let us now consider the GN model on a two-dimensional lattice,
\beq
S = N \sum_x \frac{\sigma(x)^2}{2 \lambda} + \sum_{x,y} \sum_{\alpha=1}^N \bar
\chi^\alpha(x) \left[ D_{xy} + \Sigma_{xy}\right] \chi^\alpha(y)
\eeq
where the staggered Dirac operator
\beq
D_{xy} = \frac{1}{2} \left[ \delta_{x,y+\hat 1} - \delta_{x,y-\hat 1} \right]
+ \frac{1}{2} (-1)^{x_1} \left[ e^{+\mu} \delta_{x,y+\hat 2} - e^{-\mu} \delta_{x,y-\hat 2} \right]
\eeq 
describes 2 flavours and 
\beq
\Sigma_{xy} = \frac{1}{4} \delta_{xy} \left(\sigma(x) + \sigma(x-\hat 1) +
  \sigma(x - \hat 2) + \sigma(x - \hat 1 - \hat 2)\right).
\eeq
The modification of the naive discretisation $\sigma(x) \delta_{xy} \rightarrow
\Sigma_{xy}$ is necessary to ensure the correct continuum limit
\cite{Cohen:1983nr}. With the staggered discretisation the following discrete
chiral symmetry is preserved,
\beq
\chi(x) \rightarrow (-1)^{x_1+x_2} \chi(x), \quad \bar \chi(x) \rightarrow
-(-1)^{x_1+x_2} \bar \chi(x), \quad \sigma(x) \rightarrow - \sigma(x).
\eeq
Alternatively we consider a discretisation respecting exact chiral
symmetry by employing the overlap Dirac operator
\beq
    D = m \left\{ 1 + D_W(-m) 
        \left[ D_W^\dagger(-m) D_W(-m) \right]^{-1/2} \right\}
\eeq
satisfying the Ginsparg-Wilson relation $D^\dagger + D = \frac{1}{m} D^\dagger
D$.  The coupling of the fermionic fields to the scalar field is introduced
according to
\beq 
{\cal L} = \bar \psi(x) \left[ D_{x,y} - \frac{\sigma(x)}{4 m} D_{x,y}
     - D_{x,y} \frac{\sigma(y)}{4 m}
     + \sigma(x) \delta_{x,y} \right] \psi(y) 
\eeq 
which is consistent with a scalar density transforming covariantly under a
lattice chiral symmetry transformation. For a homogeneous condensate $\sigma
\rightarrow$ const.~it is just the usual mass term of the overlap operator.

\begin{figure}[b]
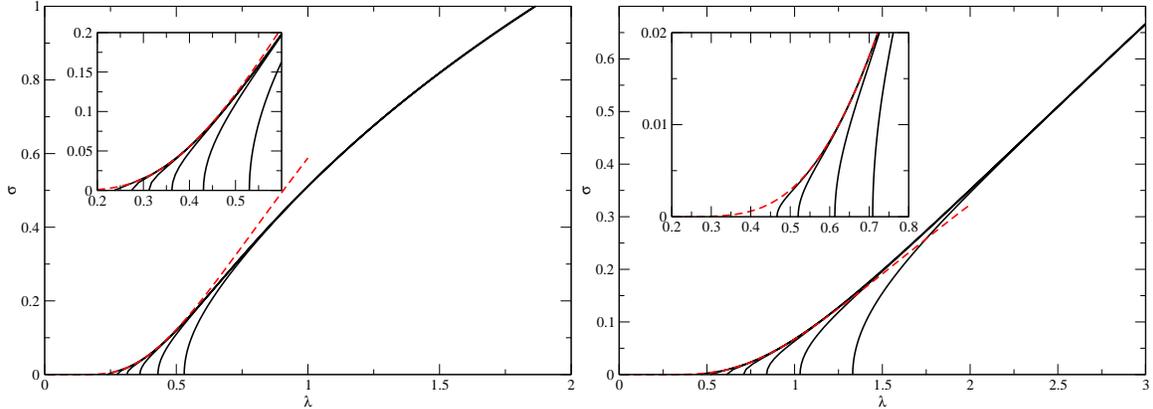

\includegraphics[width=0.5\textwidth]{Figures/fss_sigma_vs_lambda_staggered}
\includegraphics[width=0.5\textwidth]{Figures/fss_sigma_vs_lambda_overlap}
\mvcapt
\caption{
  Scaling of the condensate $\sigma$ as a function of the coupling $\lambda$
  for staggered (left) and overlap fermions (right) for various lattice sizes.
  The dashed red lines are the asymptotic scaling curves.
}
\label{fig:fss_sigma_vs_lambda}
\end{figure}

The large-$N$ limit of the lattice theory is obtained by minimising the free
energy. Using a homogeneous mean field, we obtain $\sigma$ as a function of
$\lambda$ from the the gap equation. In Figure~\ref{fig:fss_sigma_vs_lambda}
we show these scaling functions for various lattice sizes.  The dashed red
curve describes the asymptotic scaling curve $2^{3/2} e^{-\pi/2\lambda}$ for
staggered and $1.5539...  e^{-\pi/\lambda}$ for overlap fermions. As expected,
the staggered fermion formulation exhibits an additional factor of 2 in the
exponent of the scaling curve due to the doubling of the number of flavours.

\begin{figure}[t]
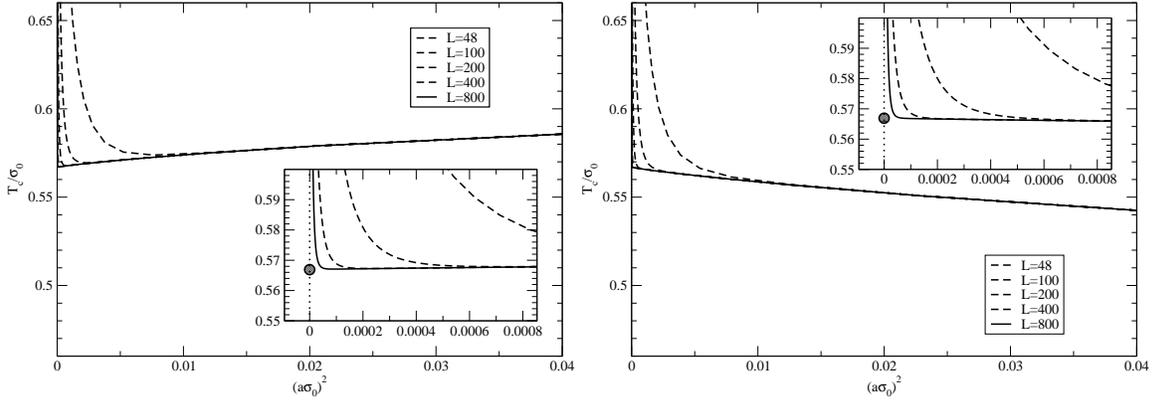

\includegraphics[width=0.5\textwidth]{Figures/scaling_Tc_over_sigma0}
\includegraphics[width=0.5\textwidth]{Figures/scaling_Tc_over_sigma0_overlap}
\mvcapt
\caption{
Scaling of
$T_c/\sigma_0$ vs $(a\sigma_0)^2$ on various lattice sizes for
the staggered (left) and the overlap operator (right).
}
\vspace{-0.2cm}
\label{fig:scaling_Tc_over_sigma0}
\end{figure}

To start our investigation of the $(\mu,T)$--phase diagram we first determine
the chiral transition temperature $T_c$ at $\mu=0$ where the chiral condensate
$\sigma(T)$ vanishes. The results are shown in
Figure~\ref{fig:scaling_Tc_over_sigma0} where we plot the scaling of
$T_c/\sigma_0$ versus $(a\sigma_0)^2$ on various lattice sizes for the
staggered operator on the left and the overlap operator on the right.  Both
formulations exhibit lattice artefacts of similar size, but with a different
sign, and nicely approach the analytically known value $T_c/\sigma_0 =
e^{C}/\pi = 0.5669...$ in the continuum \cite{Wolff:1985av}, thereby
confirming the universality of the continuum limit.

\section{Baryonic matter in the Gross-Neveu model}
\label{sec:bmGNm}
\begin{figure}[bh]
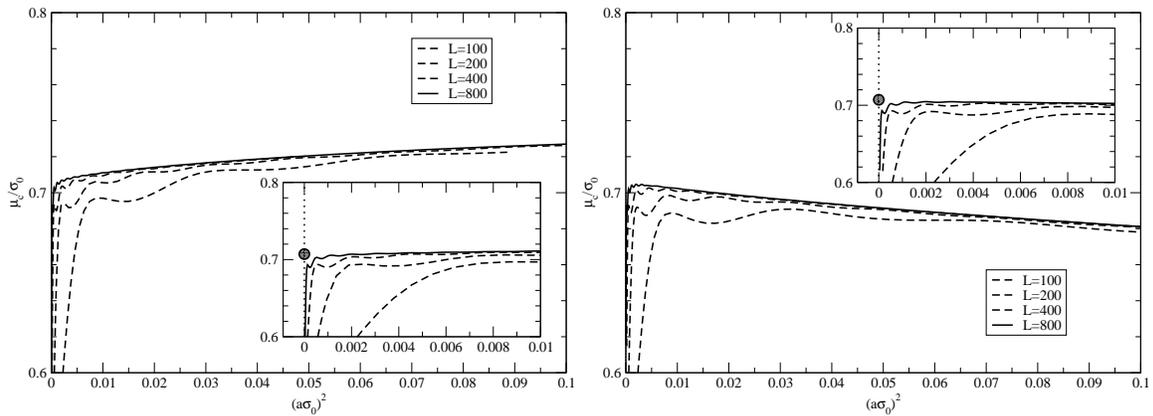

\includegraphics[width=0.5\textwidth]{Figures/scaling_muc_over_sigma0_staggered}
\includegraphics[width=0.5\textwidth]{Figures/scaling_muc_over_sigma0_overlap}
\mvcapt
\caption{
  Scaling of $\mu_c/\sigma_0$ vs $(a\sigma_0)^2$ on various lattice sizes for
  the staggered (left) and the overlap operator (right) using the homogeneous
  ansatz for the solution of the gap equation.
}
\label{fig:scaling_muc_over_sigma0}
\end{figure}

In both formulations a finite chemical potential can be introduced by
weighting the temporal derivatives with factors $\exp(\pm \mu)$
\cite{Hasenfratz:1983ba}.  We can then check the universality of the continuum
limit of $\mu_c$ at $T\simeq0$ and look at the discretisation artefacts for
the two lattice formulations. In Figure \ref{fig:scaling_muc_over_sigma0} we
show the scaling of $\mu_c/\sigma_0$ versus $(a\sigma_0)^2$ for the staggered
Dirac operator on the left and the overlap on the right, still using only the
homogeneous ansatz for the solution of the gap equation.
Again, the two formulations show discretisation artefacts of similar size, but
with a different sign, and nicely scale to the analytically known value
$\mu_c=1/\sqrt{2}$ from the homogeneous ansatz of the condensate
\cite{Wolff:1985av}. However, in both cases the continuum value is approached
non-monotonically -- this is caused by the fact that the Fermi momentum
changes continuously with $\mu$ while the lattice momentum is quantised by the
finite lattice size.  This finite volume effect would be most striking at
exactly $T=0$ where one would expect a sawtooth behaviour; at $T > 0$,
however, it is smoothened due to the softening of the Fermi surface.

\begin{figure}[h]
\begin{center}
\includegraphics[width=0.5\textwidth]{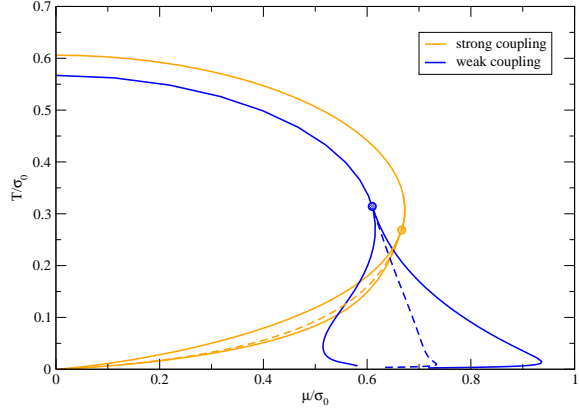}
\end{center}
\mvcapt
\caption{
  Phase diagram at weak ($L_t=80$) and strong ($L_t=4$) coupling using the
  homogeneous ansatz for the condensate. The full line above the tricritical
  point marked by the dot denotes the second order transition, the dashed line
  below denotes the first order transitions while the two full lines to the
  left and right of it bound the region of metastability associated with the
  first order transition.
}
\label{fig:mu_Tc_phase_diagram}
\end{figure}

In Figure \ref{fig:mu_Tc_phase_diagram} we show the full phase diagram for the
staggered operator at weak ($L_t=80$) and strong coupling ($L_t=4$), still
using the homogeneous ansatz for the condensate. We note that below the
tricritical point both the region of metastability and that of the chirally
broken phase shrink considerably towards strong coupling.

\begin{figure}[h]
\begin{center}
\includegraphics[width=0.5\textwidth]{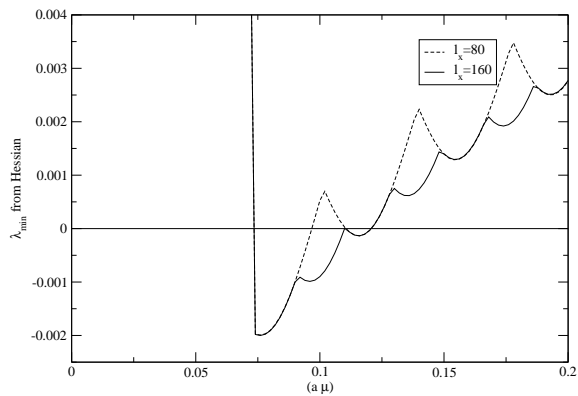}
\end{center}
\mvcapt
\caption{
  The lowest eigenvalue of the Hessian matrix associated with spatial
  variations of the condensate as a function of $\mu$ for a fixed temperature
  and two volumes.
}
\label{fig:finite_uni_cell_effect}
\end{figure}

Let us now relax the assumption of translation invariance. In order to
determine the phase boundaries related to the transitions into the new
baryonic matter phase, we perform on the one hand a brute force minimisation
of the free energy and on the other hand check for instabilities of the
homogeneous vacuum. The latter is achieved by monitoring the eigenvalues of
the Hessian matrix in the homogeneous vacuum -- a negative eigenvalue
indicates that the free energy can be lowered further by an inhomogeneous
perturbation.  This is illustrated in Figure~\ref{fig:finite_uni_cell_effect}
where we show the lowest eigenvalue of the Hessian matrix associated with
spatial variations of the condensate as a function of $\mu$ for a fixed
temperature. It is clearly seen that for $\mu \gtrsim 0.075$, where the
preferred homogeneous condensate is $\sigma = 0$, an inhomogeneous one is
favoured. In fact, a brute force free energy minimisation shows that this is
also true for some range $\mu \lesssim 0.075$, however finite-size effects
cause a small free energy barrier between the two vacua.  We also note that
the non-monotonic behaviour of the lowest eigenvalue is an artefact of the
non-commensurability of the spatial lattice size with the intrinsic length
scale of the inhomogeneous condensate (compare the two volumes in
Figure \ref{fig:finite_uni_cell_effect}).
In Figure \ref{fig:revised_phase_diagram} right we show the new phase diagram
in the thermodynamic limit, at fixed $L_t=80$, to be compared with the
continuum phase diagram on the left.  The dashed lines show the effects from
finite volume. In particular we point to the fringes at the right phase
boundary of the crystal phase which are the reflection of the
incommensurability effects discussed above. In the thermodynamic limit these
effects disappear.

\section{Summary and Outlook}
We have investigated the phase structure of the GN model in the
$(\mu,T)$--plane.  The breakdown of translational invariance of the chiral
condensate requires a revision of the GN model phase diagram. Besides the
massive and massless Fermi gas phase, a new phase of baryonic matter emerges
which forms a baryon crystal. The transition to the new phase is always second
order for any temperature. Our investigation of the phase structure on the
lattice indicates that the new crystal phase disappears at strong coupling,
because the topological excitations associated with the forming of the crystal
fall through the lattice. Furthermore, large volumes are needed to detect the
baryon crystal phase and to avoid or reduce complicated artefacts due to the
incommensurability of the intrinsic length scale of the inhomogeneous
condensate and the lattice volume, which distort both the phase boundary and
the order of the phase transition.

In this exactly solvable model, the crystal is formed by topological defects,
the kinks and antikinks, which carry the baryon number. This crystal becomes
stable for a sufficiently large chemical potential. One may wonder how general
this phenomenon is. A similar topological crystal may occur in the $(2+1)d$
Nambu-Jona-Lasinio model. If it does, perhaps it also occurs in QCD.

\bibliography{bmlGNm_proc_lattice06}

\end{document}